# Unraveling the thermodynamics and mechanism behind the lowering of reduction temperatures in oxide mixtures

Shiv Shankar[1], Barak Ratzker[1,*], Alisson Kwiatkowski da Silva[1], Tim M. Schwarz[1], Hans Brouwer[2], Baptiste Gault[1,3], Yan Ma[1,2,*], Dierk Raabe[1,*]

[1] Max Planck Institute for Sustainable Materials, Max-Planck-Str. 1, Düsseldorf 40237, Germany

[2] Department of Materials Science & Engineering, Delft University of Technology, Mekelweg 2, Delft 2628 CD, the Netherlands

[3] Department of Materials, Imperial College London, London, SW7 2AZ, UK

*Corresponding authors: b.ratzker@mpie.de (B. Ratzker), y.ma@mpie.de (Y. Ma), d.raabe@mpie.de (D. Raabe)

## Abstract

Hydrogen-based direct reduction offers a sustainable pathway to decarbonize the metal production industry. However, stable metal oxides, like $Cr_2O_3$, are notoriously difficult to reduce, requiring extremely high temperatures (above 1300 °C). Herein, we show how reducing mixed oxides can be leveraged to lower hydrogen-based reduction temperatures of stable oxides and produce alloys in a single process. Using a newly developed thermodynamic framework, we predict the precise conditions (oxygen partial pressure, temperature, and oxide composition) needed for co-reduction. We showcase this approach by reducing $Cr_2O_3$ mixed with $Fe_2O_3$ at 1100 °C, significantly lowering reduction temperatures (by ≥200 °C). Our model and post-reduction atom probe tomography analysis elucidate that the temperature-lowering effect is driven by the lower chemical activity of Cr in the metallic phase. This strategy achieves low-temperature co-reduction of mixed oxides, dramatically reducing energy consumption and $CO_2$ emissions, while unlocking transformative pathways toward sustainable alloy design.

# 1 Introduction

Metallic materials are indispensable for technological progress and industrial development[1], serving as foundational components in aerospace[2], automotive[3], energy[4,5], and construction[6] sectors. However, conventional metal production, relies on fossil-based reductants and energy-intensive pyrometallurgy, hydrometallurgy, and electrometallurgy processes to extract metals from oxide and sulfide ores[7,8]. This poses a major challenge to global decarbonization efforts[9]. The metals industry consumes approximately 10% of global energy supplies and produces enormous $CO_2$ emissions, accounting for nearly 40% of industrial greenhouse gas emissions[10]. Transitioning to sustainable metal production is thus critical to achieving carbon neutrality by 2050[11].

Hydrogen-based direct reduction (HyDR) of metal oxide mixtures offers an alternative approach towards sustainable alloy design and production[12]. The HyDR process employs hydrogen as a reducing agent, generating water vapor as a redox product[13]. However, the thermodynamic stability of certain metal oxides, like $Cr_2O_3$, necessitates prohibitively high reduction temperatures when using hydrogen (above 1100 °C). As HyDR is an endothermic process[14], the efficiency of hydrogen as a reductant is closely linked to the operating temperatures and the energy costs for reducing such stable oxides raise major technical and economical drawbacks. This well-known disadvantage motivated us to revisit HyDR of metals oxides from a more profound thermodynamic perspective, with the idea to not reduce individual oxides, like in traditional metal synthesis, but instead mixed oxides. Moreover, co-reduction of oxide mixtures offers a potential leap in sustainable manufacturing by eliminating some energy- and capital-intensive steps of conventional metallurgy, where alloys are produced through subsequent liquid-phase mixing rather than during initial metal extraction.

Several studies revealed that mixing of a stable oxide (*e.g.*, $Cr_2O_3$) with a metal or less stable oxides (*e.g.*, $Fe_2O_3$, NiO) can significantly lower reduction temperatures (by ~200 °C)[15–20] and accelerate kinetics[21,22] in both, carbothermic and hydrogen-based direct reduction. Notable examples include obtaining a Fe-Cr alloy from a $Fe_2O_3+Cr_2O_3$ oxide mixture, Co-Cr from $CoCr_2O_4$ solid solution[17], Ni-Cr from $NiO+Cr_2O_3$[18], Fe-Ni-Cr from $Fe_2O_3+NiO+Cr_2O_3$[19], Fe-Ni-Mo from $Fe_2O_3+NiO+MoO_3$, and Co-Cr-Fe-Ni from $Co_3O_4+Cr_2O_3+Fe_2O_3+NiO$[20]. Furthermore, the onset reduction temperature of $Cr_2O_3$ drops by 150-200 °C when mixed with metallic Fe[25] or $Fe_2O_3$[21].

Despite having abundant empirical evidence, only several studies attempted to explain the phenomenon of lowering reduction temperatures for stable oxides in mixtures. Barshchevskaya and Radomysel'skii[21] suggested the decrease in activity of Cr in Fe-Cr solid solution during HyDR. For instance, Zhang et al.[23] showed that co-reduction of $Fe_2O_3$+NiO lowers the apparent activation energy by 18.79 % to 71.3 kJ·mol$^{-1}$ compared with individual oxides (NiO → Ni: 87.8 kJ·mol$^{-1}$). Furthermore, Kenel et al. proposed that synergistic effects between NiO and $Fe_2O_3$ as well as the catalytic effect of Ni lead to lower reduction temperatures for the formation of Fe-Ni alloys[24]. However, the underlying thermodynamic principles and mechanisms governing the lowering of reduction temperature of stable oxides in mixtures are still unresolved.

Herein, we employed the CALculation of PHAse Diagrams (CALPHAD) approach to establish a general thermodynamic framework to assess the reducing conditions of metal oxide mixtures. By quantifying the equilibrium oxygen partial pressure ($p_{O_2}$), we predict reduction conditions and the temperatures required for the complete reduction as a function of oxide composition, using hard-to-reduce $Cr_2O_3$ as a model substance. Additionally, HyDR experiments of $Fe_2O_3$+$Cr_2O_3$ oxide mixtures, targeting Fe-10Cr and Fe-50Cr (at.%) alloys, were conducted to demonstrate the effect of temperature and composition on $Cr_2O_3$ reduction. Our findings elucidate that the lowering of the hydrogen-based reduction temperature is driven by the lower activity of Cr in Fe(Cr). The thermodynamic framework provides a universal tool for designing sustainable HyDR processes across oxide systems.

## 2 Results and discussion

### 2.1 Effect of temperature and composition on the reducibility of oxide mixtures

The reducibility of a metal oxide mixture is quantitatively represented by $p_{O_2}$, the equilibrium partial pressure of oxygen between the metal and metal oxide, calculated using the CALPHAD approach, as schematically illustrated in **Fig. 1a**. A detailed description of the equilibrium calculation steps is provided in the Methods Section. To simplify, NiO+$Cr_2O_3$ or $Fe_2O_3$+$Cr_2O_3$ oxide mixtures are hereafter defined by the target alloy composition, e.g., Fe-10Cr and Ni-10Cr oxide mixtures for the target Fe-10Cr and Ni-10Cr (at.%) alloys, respectively. The mole fraction of metals (Fe and Cr, or Ni and Cr) in their respective metallic phases (BCC or FCC) as a function of $p_{O_2}$ in Fe-25Cr and Ni-25Cr oxide mixtures is shown in **Fig. 1b** for several temperatures, i.e. 700, 900, and 1100 °C. The $p_{O_2}$ values at the onset of $Cr_2O_3$ reduction are identified as the threshold $p_{O_2}$ when metallic Cr starts to form and solutions into the Fe or Ni metallic matrix. The thermodynamic calculations indicate that the threshold $p_{O_2}$ shifts towards higher values with increasing temperature, suggesting enhanced reducibility. This result aligns with empirical findings in the literature[15]. Notably, the threshold $p_{O_2}$ for $Cr_2O_3$ is lower for the Fe-25Cr compared with that for the Ni-25Cr oxide mixture. This suggests that $Cr_2O_3$ reduction is facilitated more effectively upon mixing with NiO compared with $Fe_2O_3$.

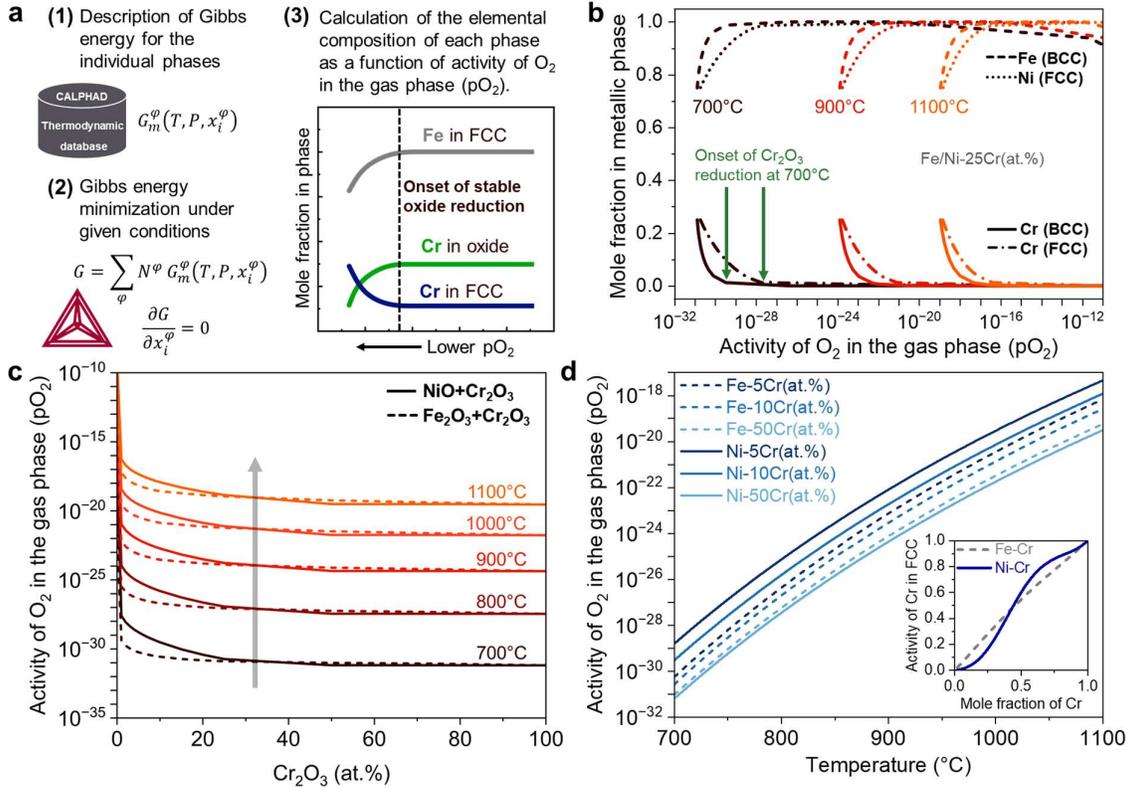

**Fig. 1. Thermodynamic stability calculations for predicting reducibility of Ni-Cr and Fe-Cr oxide mixtures.** (a) Schematic diagram showing the thermodynamic framework developed to calculate the equilibrium oxygen partial pressure ($p_{O_2}$) using the Thermo-Calc software. (b) Variation of metallic mole fraction in BCC or FCC phase with $p_{O_2}$, for determining the threshold $p_{O_2}$ at the onset of $Cr_2O_3$ reduction at 700, 900, and 1100 °C. (c) $p_{O_2}$ as a function of $Cr_2O_3$ (at.%) in $NiO+Cr_2O_3$ and $Fe_2O_3+Cr_2O_3$ oxide mixtures for the temperature range of 700-1100 °C. (d) Variation of the $p_{O_2}$ with 5, 10, and 50 at.% $Cr_2O_3$ in NiO and $Fe_2O_3$ oxide mixture in the temperature range of 700-1100 °C; the inset illustrates the change in Cr activity with the Cr content in Fe-Cr and Ni-Cr solution for FCC crystal structure at 900 °C.

The variation of $p_{O_2}$ with temperature for different $Cr_2O_3$ concentrations (at.%) in $NiO+Cr_2O_3$ and $Fe_2O_3+Cr_2O_3$ binary oxide mixtures is presented in **Fig. 1c**. The $p_{O_2}$ values for the Ni-10Cr oxide mixture increased from $3.06 \times 10^{-30}$ to $1.19 \times 10^{-18}$ as the temperature increased from 700 °C to 1100 °C. Such an increase reflects the destabilization of the oxide mixture with increasing temperature (since oxide formation is an exothermic process). Thus, oxygen removal from the metal oxide mixtures is thermodynamically more favorable at higher temperatures. A similar trend is observed for the Fe-10Cr oxide mixture, as the temperature increased from 700 °C to 1100 °C, the $p_{O_2}$ values increased from $2.75 \times 10^{-31}$ to $2.62 \times 10^{-19}$. Moreover, for both binary oxide mixtures, the effect of temperature on the $p_{O_2}$ is more pronounced at lower temperatures. At the lower temperature range (700-800 °C), an increase of temperature by 100 °C increases the $p_{O_2}$ value by at least six orders of magnitude, whereas in the high temperature range (900-1100 °C), raising the temperature by 200 °C only increases $p_{O_2}$ by up to five orders of magnitude.

Furthermore, the reduction of the oxide mixture is not only governed by the amount of stable oxide ($Cr_2O_3$) but also the type of oxide matrix (NiO or $Fe_2O_3$) in which the stable oxide is introduced (**Fig. 1d**). The $p_{O_2}$ for $Cr_2O_3$ reduction is higher for Ni-Cr oxide mixture compared with Fe-Cr oxide mixture at low $Cr_2O_3$ concentrations (≤25 at.%, **Fig. 1c**). This effect is inferred by the lower chemical activity of Cr in Ni-Cr solid solution (below ~43 at.% Cr), as shown in inset, **Fig. 1d**. The chemical activity reflects the effective concentration of a species available for reaction within a mixture, *i.e.*, it quantifies the reactivity level for a specific element in a compound. A lower chemical activity indicates greater thermodynamic stability and lower reactivity[26], consequently, the decrease in Cr activity in the metallic phase serves as the driving force for $Cr_2O_3$ reduction. Upon increasing $Cr_2O_3$ concentration, the matrix (NiO or $Fe_2O_3$, the less stable oxides) becomes less dominant, hence the decrease in chemical activity of Cr in the Fe-Cr or Ni-Cr solid solution is hindered. Therefore, the difference in $p_{O_2}$ between the Fe-Cr and Ni-Cr oxide mixtures diminishes and becomes negligible above 25 at.% $Cr_2O_3$ (**Fig. 1c**). Thus, both the concentration and the specific type of metal oxide play crucial roles in determining the overall reducing conditions. Building on these thermodynamic calculations, $Fe_2O_3+Cr_2O_3$ mixtures were chosen as the model system for HyDR experiments owing to the higher technological relevancy of Fe-Cr-based alloys.

## 2.2 Hydrogen-based direct reduction of oxide mixture samples

The mass loss of compact $Fe_2O_3+Cr_2O_3$ powder samples during HyDR was measured by thermogravimetric analysis (TGA) as a function of time and temperature (**Fig. 2**). The mass loss and corresponding reduction degrees are presented in **Table 1**. In all the experiments the onset of mass loss is observed at ~250 °C, complemented by the detection of an increase in the signal of water vapor using mass spectrometry (**Fig. S1**). Note that the moisture content in the oxide mixture powder samples was quantified to be ~1.5±0.5 wt.% by baking in an argon atmosphere at 300 °C. **Fig. 2a** presents the reduction behavior of Fe-10Cr oxide mixture at 900 °C. The observed total reduction degree (91.37%) was slightly higher than the theoretically expected value for complete reduction of $Fe_2O_3$ to Fe (89.98%) in the oxide mixture, suggesting some reduction of $Cr_2O_3$ to metallic Cr (1.31 at.%).

The mass loss curves of the Fe-10Cr and Fe-50Cr oxide mixtures at 1100 °C are shown in **Fig. 2b** and **2c**, respectively. The observed total mass loss is noticeably higher than the theoretical value predicted for the reduction of $Fe_2O_3$ to Fe (without any $Cr_2O_3$ reduction), exceeding it by 9.08% and 13.12% for Fe-10Cr and Fe-50Cr oxide mixtures, respectively. This additional mass loss suggests reduction of $Cr_2O_3$ to metallic Cr. The amount of metallic Cr in these two cases is estimated to be 3.74 and 4.40 at.%, respectively. Furthermore, the mass loss rate as a function of time exhibits an additional shallow peak during the holding period that likely corresponds to $Cr_2O_3$ reduction (**Fig. S2**).

Table 1. Thermogravimetric analysis results for Fe-Cr oxide mixtures.

| Sample | Total mass loss (mg) | Moisture content (mg) | Total reduction degree (%) | Metallic Cr (at.%) | $Cr_2O_3$ reduced (wt.%) |
|---|---|---|---|---|---|
| **Fe-10Cr 900°C** | 42.35 | 2.22 | 91.37 | 1.31 | 12.31 |
| **Fe-10Cr 1100°C** | 52.40 | 2.67 | 93.93 | 3.74 | 35.03 |
| **Fe-50Cr 1100°C** | 32.15 | 2.76 | 52.57 | 4.40 | 4.63 |
| **$Cr_2O_3$ 1100°C** | 3.28 | 3.28 | ~0 | ~0 | ~0 |

**Fig. 2d-f** shows the dual-phase microstructure, comprising a metallic and oxide phase. The Fe-10Cr sample reduced at 900 °C was more porous than that reduced at 1100 °C. In contrast, the Fe-50Cr sample was extremely porous with an abundance of unreduced $Cr_2O_3$ preventing sintering at this temperature. The corresponding energy dispersive spectroscopy (EDS) map revealed unreduced $Cr_2O_3$ (O and Cr overlap) in Fe-10Cr after reducing at 1100 °C compared with the sample reduced at 900 °C. In light of the submicron microstructure features and relatively large EDS interaction volume (≥1μm), it is not a reliable method to quantify the Cr in the metallic phase. Thus, atom probe tomography was employed to quantify the Cr content across the metallic-oxide interface (**Section 2.3**).

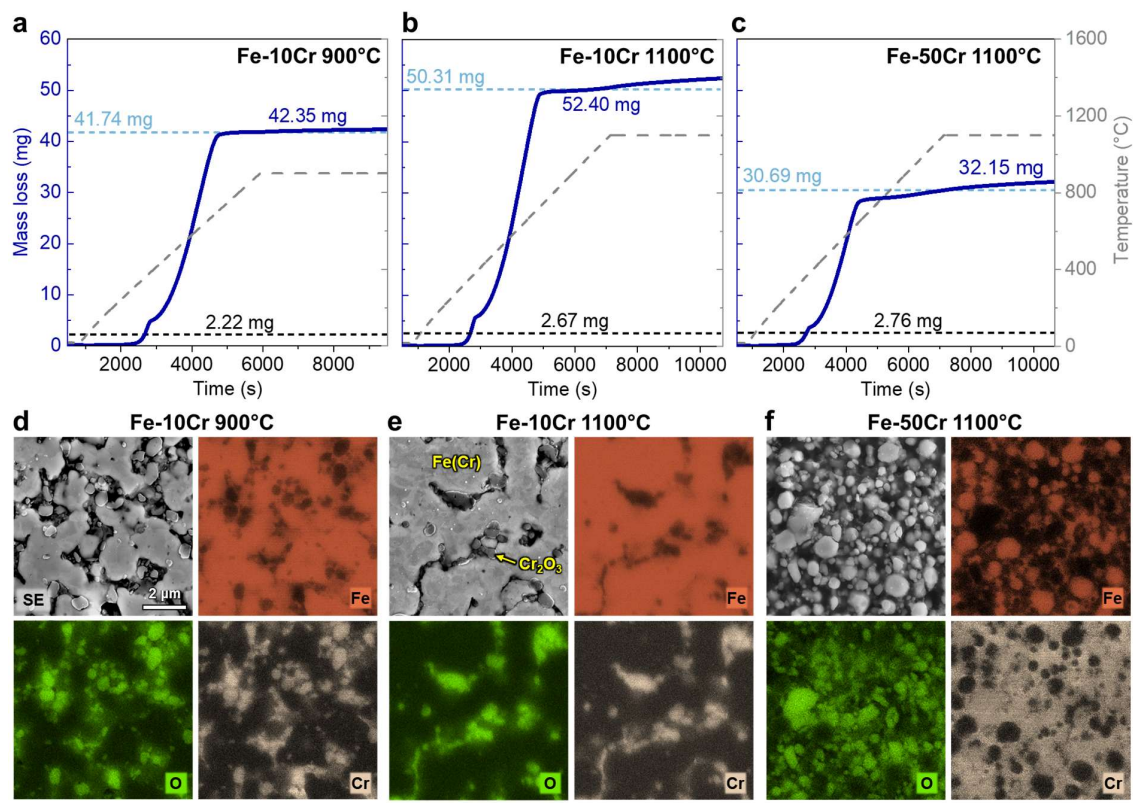

**Fig. 2. Reduction behavior and microstructure characterization of Fe-Cr oxide mixtures.** Thermogravimetric analysis results for hydrogen-based direct reduction of (a) Fe-10Cr oxide mixture at 900 °C, (b) Fe-10Cr oxide mixture at 1100 °C, and (c) Fe-50Cr oxide mixture at 1100 °C in a gas mixture of 75 vol.% $H_2$ and 25 vol.% Ar; the heating rate was 10 °C/min. The solid blue and grey dashed line represent the measured mass loss and temperature variation with time, respectively. The black dashed line represents the expected mass loss due to moisture evaporation and the light blue dashed line represents the cumulative mass loss for complete reduction of $Fe_2O_3$ to Fe including moisture evaporation. The initial mass of the Fe-10Cr 900°C, Fe-10Cr 1100°C, and Fe-50Cr 1100°C samples were 147.9, 178.3, 184.5 mg, respectively. Secondary electron (SE) micrographs and corresponding energy dispersive spectroscopy elemental maps of iron, oxygen, and chromium in the (d) Fe-10Cr (at.%) reduced at 900 °C, (e) Fe-10Cr (at.%), and (f) Fe-50Cr (at.%) oxide mixtures reduced at 1100 °C.

## 2.3 Microstructural analysis of the reduced samples

X-ray diffraction (XRD) analysis of reduced oxide mixtures is shown in **Fig. 3a** and **b** and the corresponding phase fractions are summarized in **Table 2**. The measurements revealed that the Fe-10Cr oxide mixture reduced at 900 °C comprises 89.5±0.9 wt.% Fe-Cr (BCC) and 10.5±0.4 wt.% $Cr_2O_3$. Given that the theoretical fractions of metallic Fe and $Cr_2O_3$ in the initial sample are 86.87 wt.% and 13.13 wt.% (assuming complete reduction of $Fe_2O_3$ to Fe, and no reduction of $Cr_2O_3$), these results indicated partial reduction of $Cr_2O_3$ and concur well with the TGA results. Upon increasing the reduction temperature to 1100 °C, a notable increase in BCC phase (by 5.14 wt.%) and a decrease in $Cr_2O_3$ content (by 43.80 wt.%) was observed.

Table 2. Phase fraction of reduced samples measured by XRD.

| Sample | BCC (wt.%) | $Cr_2O_3$ (wt.%) | FCC (wt.%) |
|---|---|---|---|
| **Fe-10Cr 900°C** | 89.5±0.9 | 10.5±0.4 | -- |
| **Fe-10Cr 1100°C** | 94.1±1.1 | 5.9±0.5 | -- |
| **Fe-50Cr 1100°C** | 46.2±0.5 | 48.8±0.6 | 5±0.2 |

The Fe-50Cr oxide mixture reduced at 1100 °C consisted of Fe-Cr (BCC), Fe-Cr (FCC) and $Cr_2O_3$. Since the theoretical fractions of metallic Fe and $Cr_2O_3$ are 42.37 wt.% and 57.63 wt.%, respectively, this result indicated partial reduction of $Cr_2O_3$ to metallic Cr and aligns with the TGA observations. The BCC phase is thermodynamically stable for the Fe-Cr system at lower temperatures[27,28], as shown in **Fig. S3** below 800 °C for the composition range of Cr from 1 to 100 at.%. The presence of FCC phase in the reduced Fe-50Cr sample at room temperature may result from local stabilization effects[29].

In addition, the reduction of $Cr_2O_3$ to metallic Cr is inferred from the change in the lattice parameter observed for the BCC phase (lattice expansion) due to the substitutional dissolution of Cr. As shown in **Fig. 3a**, the BCC (211) peak shifted to a lower 2θ angle from 99.7° (Fe-10Cr 900°C) to 99.6° (Fe-10Cr 1100°C), indicating an increase in the lattice parameter. The BCC lattice parameter was calculated to be 2.867 Å for the Fe-10Cr (at.%) oxide mixture reduced at 900 °C. With an increase in reduction temperature to 1100 °C, the lattice parameter increased to 2.869 Å and further to 2.870 Å for the Fe-50Cr (at.%) oxide mixture reduced at 1100 °C, as shown in **Fig. 3b**. Additionally, an increase in the $Cr_2O_3$ lattice parameter was also observed. The *a* and *b* lattice parameter increased from 4.955 Å for Fe-10Cr 900 °C to 4.960 Å for Fe-50Cr1100 °C. This suggests some dissolution of Fe into the $Cr_2O_3$ lattice, corresponding to ~0.5 wt.% $Fe_2O_3$ in $Cr_2O_3$[30].

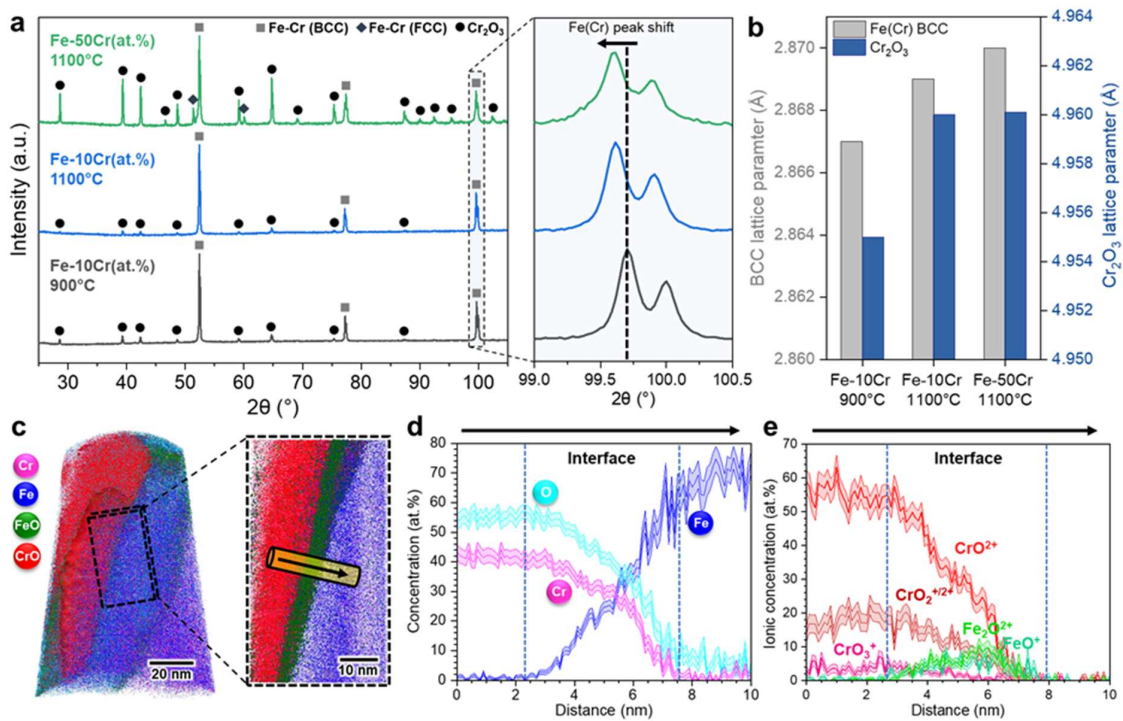

**Fig. 3. Structural and chemical analysis of reduced Fe-Cr oxide mixtures.** (a) X-ray diffractograms of Fe-10 Cr (at.%) oxide mixtures reduced at 900 °C and 1100 °C, and Fe-50Cr (at.%) mixed oxide sample reduced at 1100 °C, with a heating rate of 10 °C/min; the blown-up region shows the peak shift caused by the expansion of BCC lattice due to Cr dissolution. (b) Lattice parameters of the BCC metallic phase and $Cr_2O_3$ after the HyDR of Fe-Cr oxide mixtures at 900 and 1100 °C. (c) APT investigation of Fe-10Cr (at.%) oxide mixture sample reduced at 1100 °C; 3D reconstruction of the measured volume from the Fe-$Cr_2O_3$ interface. Compositional profile in (d) atomic and (e) ionic fraction along the oxide/metal interface, indicating mutual dissolution of Cr into Fe and Fe into $Cr_2O_3$.

To understand the elemental partitioning between $Cr_2O_3$ and Fe, we performed atom probe tomography (APT) analysis of Fe-10Cr oxide mixture reduced at 1100 °C, **Fig. 3c-e**. The 3D reconstructed volume reveals a distinct interface between the oxide and metallic regions, shown in **Fig. 3c**. The compositional profile was used to analyze the chemical partitioning across the interface (marked by black arrow), either by the distribution of detected ions (**Fig. 3e**) or by an atomic distribution where all detected ions were decomposed into their constituent elements (**Fig. 3d**). It should be noted that the specific fractions of the ions detected in oxide containing APT specimens do not readily mirror the exact stoichiometry and structure of the phases at such hetero-interfaces. Due to complex evaporation mechanisms, especially when probing oxides using laser-assisted APT, the formation of complex ions during APT measurements is possible[31,32].

The compositional profile (**Fig. 3d**) shows a gradual transition from the oxide to the metallic phase, where the concentrations of Cr and O decrease whilst the Fe concentration increases.

The interface oxide region contains Fe, possibly suggesting the dissolution of Fe cations into the oxide phase during reduction. As the interface was parallel to the tip axis, it alleviates concerns usually arising by the high electrostatic field applied to the specimen, associated with inward diffusion of oxygen in a layered metal/oxide system[33]. Furthermore, the formation of oxide complexions at the metal/oxide interface in similar material systems (Fe/Fe$_3$O$_4$) has been observed, with the interfacial region predicted to exhibit higher thermodynamic stability compared with the interface devoid of these complexions[34]. It is known that residual hydrogen in the measurement chamber can lead to a higher hydrogen content being detected in the APT measurement[35], therefore we cannot state whether the detected hydrogen content (**Table 3**) originates from the HyDR process or from the APT measurements.

**Table 3.** Concentrations of elements determined by APT in regions near the oxide/metal interface and in the metallic phase far away from the interface (in at.%).

| Region | Fe | Cr | O | H |
|---|---|---|---|---|
| Oxide near the interface | 0.73±0.112 | 42.27 ± 0.893 | 55.18 ± 0.893 | 1.81 ± 0.185 |
| Metallic phase near the interface | 94.27±0.395 | 3.24 ± 0.174 | 0.35 ± 0.112 | 2.11 ± 0.299 |
| Metallic phase ~1 µm away from the interface | 93.67±0.011 | 4.24 ± 0.009 | 0.03 ± 0.001 | 2.05 ± 0.004 |

## 2.4 Mechanisms of mixed oxide reduction

### 2.4.1 Reduction mechanism at the metal-oxide interface

A schematic representation of the reduction mechanisms at the interface is presented in **Fig. 4a-c**, illustrating mutual dissolution of Cr and Fe into the opposing metal and oxide phases. The reduction of Cr$_2$O$_3$ adjacent to metallic Fe at lower temperatures is facilitated by the dissolution of Cr in metallic Fe(Cr) solid solution, resulting in lower chemical activity of Cr in the system. Therefore, reduction takes place at the interface of Cr$_2$O$_3$ and Fe particles where there is access for inbound H$_2$ and outbound H$_2$O gases (**Fig. 4a**). The interface between the metal and oxide phases comprises a transition region that may involve oxide complexions (**Fig. 4b**). Whereby, the reduction is accompanied by mutual dissolution of Cr and Fe across the interface into the opposing metallic and oxide phases, respectively, as schematically depicted in **Fig. 4c**. The thermodynamic driving force for this dissolution is lowering chemical activity of Cr in metallic Fe(Cr) and Fe in the oxide phase. The structure and precise composition of the conceivable interfacial Fe-rich oxide phase cannot be determined using APT. Nevertheless, the elemental concentration measured by APT reveals up to ~35 at.% of Fe at the oxide/metal interface (**Fig. 3d** and **e**), which then decreases over the distance of ~4 nm to a stable value of 0.73±0.11 at.% in $Cr_2O_3$ (**Table 3**). We note that the XRD analysis shows no change in the $Cr_2O_3$ lattice parameter after reducing at 900 °C (**Fig. 3b**), indicating that Fe is introduced as a consequence of the $Cr_2O_3$ reduction process.

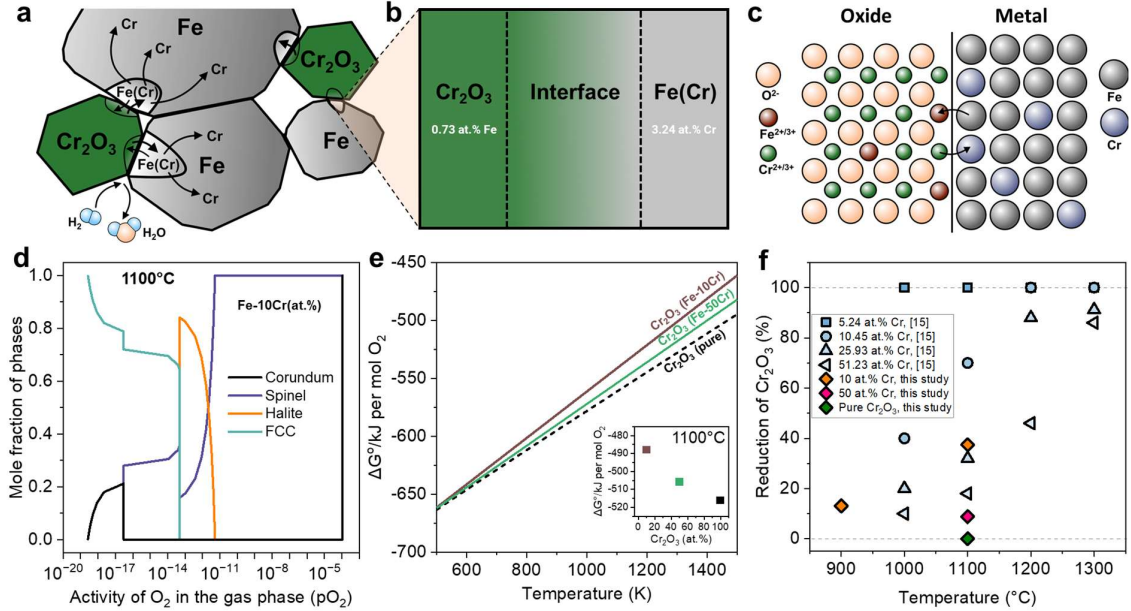

**Fig. 4. Mechanism and thermodynamic assessment of Fe-Cr oxide mixtures.** Schematic illustrations of the co-reduction mechanisms at the (a) microscopic, (b) nanoscopic, and (c) atomic scales, representing mutual dissolution of Cr and Fe into Fe(Cr) and $Cr_2O_3$, respectively. (d) Phase evolution with equilibrium partial pressure of oxygen ($p_{O_2}$) for Fe-10Cr oxide mixture at 1100 °C calculated using Thermo-Calc. (e) Multicomponent-Ellingham diagram for Fe-10Cr (at.%) and Fe-50Cr (at.%) alloys (brown and green solid lines, respectively) compared with the stability of pure $Cr_2O_3$ (black dashed line). $\Delta G^0$ for the oxidation of Fe-Cr alloy were calculated upon substituting the $p_{O_2}$ values at 973, 1073, 1173, 1273, and 1373 K obtained from thermodynamic calculations using Thermo-Calc. (f) $Cr_2O_3$ reduction degree as a function of temperature for varying composition of Cr (at.%) in the oxide mixture; comparison of the experimental results from this study with data from the literature[15].

In the metallic region, ~10 nm away from the oxide/metal interface, 3.24±0.17 at.% Cr was detected (**Table 3**). Moreover, 4.24±0.01 at.% Cr was found to be homogenously distributed throughout the volume of the metallic phase, by measuring another APT tip that did not contain oxide, prepared from a region ~1 μm away from the interface (**Fig. S6**). These findings validate the TGA results of ~3.74 at.% metallic Cr (**Fig. 2b**). These results also concur with the respective increase in the lattice parameters of metallic and oxide phases measured by XRD (**Fig. 3b**). To summarize, $Cr_2O_3$ reduction at lower temperatures is enabled by the dissolution of Cr into Fe, driven by lowering its chemical activity. The metallic Cr is then diffused away from the interface into the Fe bulk, enabling further Cr to dissolve and continuous reduction of $Cr_2O_3$ until reaching an equilibrium between the chemical activity of Cr in the oxide and metallic phase.

The phase evolution of the Fe-10Cr oxide mixture at 1100 °C with $p_{O_2}$, calculated using Thermo-Calc, is presented in **Fig. 4d**. At the final equilibrium state ($p_{O_2}$ = 2.90 × 10$^{-17}$), only FCC and corundum phases co-exist. Further reduction in $p_{O_2}$ results in the reduction of $Cr_2O_3$ to Cr, which dissolves into the Fe lattice, forming an Fe-Cr solid solution. However, these thermodynamic calculations cannot predict non-equilibrium reactions such as interfacial complex oxide phases that can form during reduction as observed by APT.

### 2.4.2 *Thermodynamic stability of mixed oxides: introduction of multicomponent-Ellingham diagrams*

The Ellingham diagram is widely used in process metallurgy to assess the reduction of single-component oxides[36]. However, the thermodynamic understanding behind the reduction of multicomponent oxide systems, or even binary oxide mixtures, remains unexplored. With the aid of thermodynamic calculations, the $p_{O_2}$ required for the complete reduction of oxide mixtures can be used to construct an Ellingham diagram for oxide mixtures, which we term as a multicomponent-Ellingham diagram. A similar approach to modify the Ellingham diagram was recently applied by Epifano and Monceau[37] to predict the oxidation of various alloys.

Herein, we constructed a multicomponent-Ellingham diagram for Fe-Cr oxide mixture (**Fig. 4e**). The diagram shows that pure $Cr_2O_3$ is highly stable, and it is placed below Fe-Cr oxide mixtures. In contrast, Fe-10Cr and Fe-50Cr are less stable and appear above $Cr_2O_3$ with their slope increasing with Fe concentration. This indicates that mixing $Cr_2O_3$ and $Fe_2O_3$ destabilizes $Cr_2O_3$, resulting in possible reduction of $Cr_2O_3$ in mixture at higher $p_{O_2}$ than that for pure $Cr_2O_3$. The destabilization effect is more pronounced in the Fe-10Cr oxide mixture compared with the Fe-50Cr oxide mixture due to the presence of a larger amount of Fe in the matrix available for the dissolution of Cr, as postulated by the co-reduction mechanism depicted in **Fig. 4a-c**. The difference in stability between the oxide mixtures and pure $Cr_2O_3$ becomes more distinct at higher temperatures (>800 °C). This could be explained by the direct relationship of the Gibbs energy with temperature (**Eq. 4**). The Gibbs energies of the studied oxide mixtures and pure $Cr_2O_3$ at 1100 °C are presented in **Fig. 4e** inset.

The equilibrium partial pressure of oxygen ($p_{O_2}$) at a given temperature can be used to predict the complete reduction of metal oxide mixtures. When using hydrogen as a reductant, $p_{O_2}$ is correlated to $p_{H_2}/p_{H_2O}$ ratio according to:

$$ln\,(p_{O_2}) = \frac{\Delta G^o}{RT} - 2ln\frac{p_{H_2}}{p_{H_2O}} \qquad (1)$$

Where, $\Delta G^o$ is the standard Gibbs free energy (298 K, 1 atm), R is the gas constant, $T$ is the temperature in Kelvin, and $p_{H_2}/p_{H_2O}$ is the ratio of the partial pressure of hydrogen and water vapor. The reduction threshold estimated using Thermo-Calc was in the order of $p_{O_2}$ ~$10^{-22}$ (**Fig. 1c**), as complete reduction of $Fe_2O_3$ to metallic iron is achieved at 700 °C (for similar TGA systems, heating rates and holding time as employed in this study)[38]. For the Fe-10Cr oxide mixture at 900 °C and 1100 °C, and the Fe-50Cr oxide mixture at 1100 °C, the corresponding $p_{O_2}$ values were determined to be 2.90×$10^{-24}$, 2.62×$10^{-19}$, and 5.76×$10^{-20}$, respectively (**Fig. 1c**).

The comparison of the predicted $p_{O_2}$ values with the estimated $p'$ ($10^{-22}$) indicates that the complete reduction of the Fe-10Cr oxide mixture at 900 °C is thermodynamically not feasible ($p_{O_2} < p'$), whereas complete reduction for both Fe-10Cr and Fe-50Cr oxide mixtures at 1100 °C should be feasible as $p_{O_2} > p'$. The experimental results generally agree with these predictions (**Fig. 4f**): for the Fe-10Cr oxide mixture reduced at 900 °C, $Fe_2O_3$ was completely reduced to Fe, while a small fraction (12.31 wt.%) of the initial amount of $Cr_2O_3$ reduced to Cr. The fact that some $Cr_2O_3$ reduced may be attributed to the presence of lower $p'$ (<$10^{-22}$) in the system, which facilitated the reduction of $Cr_2O_3$ at 900 °C. For the Fe-10Cr, and Fe-50Cr oxide mixtures reduced at 1100 °C, 35.03 wt.% and 4.63 wt.% $Cr_2O_3$ was reduced to Cr, respectively. The incomplete reduction of $Cr_2O_3$ could also be attributed to the sintering of metallic Fe particles at high temperatures, kinetically limiting the access of hydrogen gas to the $Cr_2O_3$ particle and moreover the outward transport of water vapor[39]. Additionally, since the sample was isothermally held for only 1 h at 1100 °C, longer holding times should result in further reduction of $Cr_2O_3$, similar to the findings of Barshchevskaya et al.[21].

A comparison was drawn between the reduction of Fe-Cr oxide mixtures determined empirically in the literature by Nadler et al.[15] and in this study. The effect of temperature and composition on the HyDR of mixed oxides follows the trend predicted by thermodynamic calculations (**Fig. 1c and d**), wherein higher temperatures are required to reduce mixtures with a larger fraction of $Cr_2O_3$. As shown in **Fig. 4f**, partial reduction was observed for Fe-10.45Cr and Fe-51.23Cr (at.%) oxide mixtures reduced at 1100 °C, aligning with our results. The Fe-10.45Cr (at.%) oxide mixture was fully reduced at temperatures above 1100 °C, while the Fe-25.93Cr (at.%) oxide mixture required temperatures above 1200 °C[15]. The empirical observations agree with the thermodynamic predictions, whereby $p_{O_2}$ decreases with increasing $Cr_2O_3$ content, inhibiting the reduction process.

In summary, this study elucidates how reducing mixed metal oxides with hydrogen—rather than individual oxides—lowers reduction temperatures of stable oxides (*e.g.*, $Cr_2O_3$). We developed a robust thermodynamic framework to quantify the driving force and reducibility of oxide mixtures as a function of composition and reduction temperature. The correspondingly derived multicomponent-Ellingham diagram approach serves as a practical tool to guide both alloy design and process optimization. As a proof of concept, hydrogen-based direct reduction experiments on $Fe_2O_3$+$Cr_2O_3$ mixtures targeting Fe-10Cr and Fe-50Cr (at.%) alloys showed the effect of temperature and composition on the reducibility of oxide mixtures. XRD and APT analysis of the Fe-10Cr system reduced at 1100 °C were used to confirm the presence of Cr in the metallic phase (~4.2 at.%) and revealed that also some Fe (~0.7 at.%) dissolved into the oxide phase. The enhanced reducibility of $Cr_2O_3$ at lower temperatures is mechanistically driven by dissolution of Cr into Fe, as the formation of Fe(Cr) solid solution lowers the chemical activity of Cr in the system and stabilizes Cr in the metallic phase. These experimental findings strongly support our thermodynamic calculations and underscore the synergistic interplay between composition and temperature in mixed oxide reduction. While the thermodynamic framework provides essential guidelines for selecting oxide compositions and reduction conditions, it is important to note its limitations: kinetic and morphological factors (e.g., heating rates, particles sizes, porosity, local diffusion barriers) are not accounted for the current model and may influence practical co-reduction process. Nevertheless, this study establishes a foundational tool for a unified strategy toward targeted alloy design and process optimization via hydrogen-based direct reduction, paving the way for more energy-efficient and $CO_2$-lean metal production.

## 3 Methods

### 3.1 Thermodynamic calculations

The thermodynamic stability of a metal oxide is given by the Gibbs free energy of the oxidation reaction of 1 mol of metal with oxygen (**Eq. 1**):

$$M_{(s)} + O_{2(g)} \rightarrow MO_{2(s)} \qquad \textbf{Eq. 1}$$

Where M and $MO_2$ are metal and metal oxide, respectively.

The Gibbs energy ($\Delta G$) of the oxidation reaction is calculated by:

$$\Delta G = G_{MO_2} - G_{O_2} - G_M \qquad \textbf{Eq. 2}$$

Where $G_{MO_2}$, $G_{O_2}$, and $G_M$ are the Gibbs energy of the metal oxide, oxygen, and the metal respectively [40].

Using the equation; $G = G^O + RT \ln(\frac{p}{p^0})$, **Eq. 2** can be modified to **Eq. 3**:

Where $G^O$ is the standard free energy at the standard pressure, $p^O$. The standard pressure for such reactions is assumed to be 1 atm. Thus the free energy change of reaction is determined by the relative quantities of reactants and products[41].

$$\Delta G = \Delta G^O + RT \ln Q \qquad \textbf{Eq. 3}$$

Where $\Delta G$ is the Gibbs energy change for the oxidation reaction, $\Delta G^O$ is the standard Gibbs free energy (298 K, 1 atm), R is the gas constant, and $Q$ is the reaction quotient (relative quantities of reactant and products).

At equilibrium, $\Delta G$ for the chemical reaction is zero, and the standard $\Delta G^O$ for the metal oxidation reaction is given by **Eq. 4**, assuming the activity of metal and metal oxides are unity in the solid state[36].

$$\Delta G^0 = -RT \ln K = RT \ln(p_{O_2}) \qquad \textbf{Eq. 4}$$

Where $K$ is the equilibrium constant, and $p_{O_2}$ is the equilibrium partial pressure of oxygen between the metal and metal oxide.

For binary (and multicomponent) oxide mixtures, a thermodynamic framework was established to predict the complete reducibility of a metal oxide in oxide mixtures based on the CALPHAD approach[42] using Thermo-Calc software version 2024a, **Fig. 1a**. The thermodynamic calculations of the two ternary Ni-Cr-O and Fe-Cr-O systems were performed using the metal oxide solutions TCOX10 database. All oxide and metallic phases are described as solution phases in the TCOX10 database and the equilibrium is computed by the Gibbs energy minimization of the system. The equilibrium partial pressure ($p_{O_2}$) is the activity of $O_2$ in the gas phase (with pure $O_2$ at the same temperature as the reference state) which equilibrates the oxygen chemical potential of the system. The reducibility of an oxide mixture was predicted upon comparing the partial pressure of oxygen in the system ($p'$) and the equilibrium partial pressure of oxygen ($p_{O_2}$) obtained when metal and oxide coexistes. If $p'$ for an oxide mixture exceeds the equilibrium $p_{O_2}$, the metal oxide state is stable and the reduction is not feasible. Conversely, if $p'$ is lower than equilibrium $p_{O_2}$, the alloy state is stable, and reduction is feasible. Based on the selected systems (*i.e.*, Ni-Cr-O and Fe-Cr-O), the total number of moles of metal and oxygen atoms for the given composition were introduced into the equilibrium at a fixed temperature and gas pressure. For example, a mixture of NiO (90 at. %) and $Cr_2O_3$ (10 at. %) corresponds to about 0.9 mol Ni, 0.1 mol Cr, and 1.05 mol oxygen atoms. The activity of oxygen in the gas phase (assuming ideal gas behavior) is modeled by varying the amount of oxygen in the equilibrium from zero (metallic state) to the total number of moles of oxygen in the metal oxide mixtures. $Cr_2O_3$ was chosen as the stable oxide one model substance for thermodynamic calculations because its high stability makes it highly sensitive to $p_{O_2}$ with varying compositions of the binary oxide mixtures. The effects of temperature, composition, and the metal oxide type on the reducibility of the binary oxide mixtures of NiO+$Cr_2O_3$, and $Fe_2O_3$+$Cr_2O_3$ were investigated. The equilibrium $p_{O_2}$ for the binary oxide mixtures was calculated at 700, 800, 900, 1000, and 1100 °C, respectively, with the composition of the binary oxide mixture varied from 0 to 100 at.% of the metal oxide ($Cr_2O_3$).

## 3.2 Oxide powder mixture preparation

Two different metal oxide mixtures consisting of $Fe_2O_3$ and $Cr_2O_3$ were mixed with a targeted composition (after complete co-reduction) of Fe-10Cr and Fe-50Cr in at.%. A powder mixture of about 14 g was weighed for each composition, as shown in **Table 4.** The powders were weighed using a weighing balance having an accuracy of ±0.0001 g. The measured oxide powders were mixed and homogenized using a planetary ball mill (Fritsch 7), using stainless steel balls as grinding media. Ball milling was performed at 250 rpm for 15 cycles, with each cycle lasting for 20 min and a pause for 5 min between subsequent cycles. After each cycle, the direction of rotation was reversed. Thereafter, the powder mixture was thoroughly removed from the ball-mill crucible using ethanol. The ethanol was dried in an oven at 105 °C for 1 h. The dried powder mixture was scraped off and collected.

Table 4. Composition of metal oxides in the $Fe_2O_3+Cr_2O_3$ powder mixtures.

| Metal oxide | Targeted metal content in mixture (at.%) | Metal in oxide (wt.%) | Targeted metal content in mixture (wt.%) | Weight of oxides (g) | Wt.% of oxides (%) |
|---|---|---|---|---|---|
| **Composition 1: Fe-10Cr alloy (at.%)** | | | | | |
| $Fe_2O_3$ | 90 | 70.00 | 90.62 | 12.95 | 90.43 |
| $Cr_2O_3$ | 10 | 68.43 | 9.37 | 1.37 | 9.57 |
| Total | 100 | -- | 100 | 14.32 | 100 |
| **Composition 2: Fe-50Cr alloy (at.%)** | | | | | |
| $Fe_2O_3$ | 50 | 70.00 | 51.78 | 7.40 | 51.23 |
| $Cr_2O_3$ | 50 | 68.43 | 48.22 | 7.04 | 48.77 |
| Total | 100 | -- | 100 | 14.44 | 100 |

### 3.3 Cold compaction

For each sample a powder mixture of ~1.5 g was poured into a 13 mm diameter tungsten carbide mold and compacted using a hydraulic press at a force of 30 kN. After pressing, a disc-shaped green body compact was obtained.

### 3.4 Hydrogen-based direct reduction

Non-isothermal reduction tests were carried out on ~200 mg samples cut out of the green body compacts prepared. The experiments were conducted using a thermogravimetric analysis (TGA) system (THEMYS DUO, Setaram) with an accuracy of 0.01 μg. The sample was placed in an alumina crucible with a dimension of 11 mm inner diameter and 20 mm height. Before heating the furnace, the reaction chamber was flushed with Ar gas at a flow rate of 50 ml/min for 10 min. After that the hydrogen gas with a purify of 99.999% was introduced at a flow rate of 150 ml/min. The samples were heated to 900 and 1100 °C with a heating rate of 10 °C/min, and held isothermally for 1 h at 900 and 1100 °C. The mass change of the sample was recorded by TGA at an interval of every second. After the completion of the reduction reaction, the TG furnace was switched off and the sample was cooled to room temperature in the furnace at 10 °C/min. The hydrogen flow was maintained until the furnace reached room temperature, to avoid any oxidation of the reduced sample during the cooling period.

### 3.5 Materials characterizations

The phase analysis before and after the reduction of samples was carried out by X-ray diffraction using a D8 Advance A25-X1 diffractometer, equipped with a Co Kα radiation (λ = 1.78897 Å), operated at 35 kV and 40 mA. The scanning range 2θ was from 20° to 130° with a scanning step of 0.03°. The data post-processing and plotting were conducted using Origin Pro 2022 software. The phases were determined according to the PDF-4+ database[43]. A quantitative phase analysis was performed by a pseudo-Voight function Rietveld refinement using the MDI JADE 10 software package.

The grain morphology and local chemical composition of samples were characterized using a ZEISS Sigma microscope equipped with an EDAX APEX Advanced Energy Dispersive Spectroscopy (EDS) detector. The acceleration voltage and beam current were 15 kV and 9.5 nA, respectively.

### 3.6 Atom probe tomography

Needle-shaped APT tip samples were prepared from the reduced Fe-10Cr using a Ga-FIB (Helios 5 CX), following a standard sample preparation protocol as described by Thompson *et al.*[44]. APT experiments were performed using a Cameca LEAP 5000 XR (reflectron) atom probe. A reflectron instrument was used to achieve higher mass resolution, to unambiguously distinguish the isotopes of Cr and Fe, and to minimize peak overlap. The APT measurement was performed in laser pulse mode with a laser energy of 40-50 pJ and a repetition rate of 125 kHz at a sample base temperature of 50 K and a detection rate of 0.005 ions per pulse. Data reconstruction was performed using Cameca's Integrated Visualization and Analysis Software (IVAS) in AP Suite 6.3.2. Due to the overlap of the signals of $^{54}Cr^{2+}$ and $^{54}Fe^{2+}$ at a mass-to-charge state ratio of 27 Da, peak convolution was performed to accurately determine the composition. Therefore, the main signals of $^{52}Cr$ and $^{56}Fe$ were used to calculate the isotope ratios of all isotopes based on the abundance of the main signal. Therefore, the amount of Fe and Cr ions in the 27Da signal can be determined. Two tips were analyzed, one comprising both the oxide and metallic phase to study the interface and a second metal-only specimen to validate the chemical homogeneity of metallic Cr in Fe(Cr).

## 5 Acknowledgments

S.S. acknowledges the financial support from Horizon Europe project HAlMan co-funded by the European Union grant agreement (ID 101091936). B.R. is grateful for the financial support of a Minerva Stiftung Fellowship and Alexander von Humboldt Fellowship (Hosted by D.R.). T.M.S. gratefully acknowledges the financial support of the Walter Benjamin Program of the German Research Foundation (DFG) (Project No. 551061178). T.M.S. and B.G. are grateful for funding from the DFG through the award of the Leibniz Prize 2020 from B.G.. Y.M. acknowledges financial support from Horizon Europe project HAlMan co-funded by the European Union grant agreement (ID 101091936) and the Walter Benjamin Programme of the Deutsche Forschungsgemeinschaft (Project No. 468209039). D.R. is grateful for financial support from the European Union through the ERC Advanced grant ROC (Grant Agreement No. 101054368). Views and opinions expressed are however those of the author(s) only and do not necessarily reflect those of the European Union or the ERC. Neither the European Union nor the granting authority can be held responsible for them. S.S. extends appreciation to Martina Ruffino, Rebeca Miyar, and Anurag Bajpai for insightful discussions. Uwe Tezins, Andreas Sturm, and Christian Broß are acknowledged for their support to the FIB & APT facilities at MPI SusMat. We thank Benjamin Breitbach for the XRD measurements.

## 6 Author contributions

S.S. conceptualized the study, performed thermodynamic calculations and experimental work, and wrote the manuscript. B.R. conceptualized the study, assisted with data analysis, reviewed and edited the manuscript. A.K. assisted with thermodynamic calculations, reviewed and edited the manuscript. T.M.S. performed the APT measurements, reviewed and edited the manuscript. H.B. performed TGA experiments. B.G. supervised APT experiments, reviewed and edited the manuscript. Y.M conceptualized the study, supervised the work, acquired the funding, reviewed and edited the manuscript. D.R. supervised the work, acquired the funding, reviewed and edited the manuscript.

## 7 Competing interests

The authors declare that they have no known competing financial interest or personal relationships that could have appeared to influence the work reported in this paper.

## 8 Supplementary information

Supplementary information is available for this paper.

## 9 Data availability

All data needed to evaluate the conclusions in the paper are included in the paper and/or supplementary materials. Additional data for this work can be requested from the contributed authors.